\documentclass[aps,pra]{revtex4}

\usepackage{amsmath,amssymb,bbm}
\usepackage{amsfonts}
\usepackage{graphicx}
\usepackage{pgf}

\newcommand{\omG}{\omega_G}
\newcommand{\si}{\sigma}
\newcommand{\nucl}{\mathrm{nucl}}

\newcommand\ro{\hat\rho}
\newcommand\Ho{\hat H}
\newcommand\rb{\mathbf{r}}
\newcommand\rbp{\mathbf{r^\prime}}

\newcommand\xba{\mathbf{\overline x}}
\newcommand\xbh{\mathbf{\hat x}}
\newcommand\Xbh{\mathbf{\hat X}}

\newcommand\pbh{\mathbf{\hat p}}
\newcommand\Pbh{\mathbf{\hat P}}

\newcommand\ubh{\mathbf{\hat u}}
\newcommand\pib{\boldsymbol{\pi}}
\newcommand\pibh{\mathbf{\hat\pib}}
\newcommand\fa{f^0}
\newcommand\fh{\hat f}
\newcommand\Dcal{\mathcal{D}}
\newcommand\dd{\mathrm{d}}
\newcommand\kb{\mathbf{k}}
\newcommand\lb{\mathbf{l}}
\newcommand\erm{\mathrm{e}}
\newcommand\dg{^\dagger}
\newcommand\im{\mathrm{i}}

\begin{document}

\title{Gravity-related spontaneous collapse in bulk matter}
\author{Lajos Di\'osi}
\affiliation{Wigner Research Center for Physics\\ 
             H-1525 Budapest 114, P.O.Box 49, Hungary}

\date{April 25, 2014}

\begin{abstract}
In the DP-model, gravity-related spontaneous wave function collapses suppress undesirable
Schr\"odinger Cat states. We derive the equations of the model for the
hydrodynamic-elastic (acoustic) modes in a bulk. Two particular features are discussed:
the universal dominance of spontaneous collapses at large wavelengths, and the
reduction of spontaneous heating by a slight refinement of the DP-model.
\end{abstract}

\maketitle

\section{Introduction}
After Schr\"odinger's famous thought experiment, superpositions of 
macroscopically different quantum  states are called Schr\"odinger Cat
states (or Cats, simply). Their existence in Nature would be problematic, 
particularly for our concept of gravitation and space-time. A gentle
modification of the superposition principle might suppress Cats.
Consider a massive system in a quantum state $\vert f_1\rangle$ of well-defined 
spatial mass distribution $f_1$, and consider an other state $\vert f_2\rangle$ as well.
If $f_1$ and $f_2$ are `macroscopically' different, the superposition 
\begin{equation}\label{Cat}
\frac{\vert f_1\rangle+\vert f_2\rangle}{\sqrt{2}} 
\end{equation}
represents a Cat. We quantify the measure of  'catness' as 
\begin{equation}\label{catness}
\ell_G^2=-U_{11}-U_{22}+2U_{12},
\end{equation}
where $U_{ij}$ are the formal Newton interaction potentials between 
the mass distributions $f_i,f_j$, for $i,j=1,2$ in turn. 
A spontaneous collapse of the Cat (\ref{Cat}) is then postulated:
\begin{equation}
\frac{\vert f_1\rangle+\vert f_2\rangle}{\sqrt{2}}\Longrightarrow
\begin{array}{cl}\mathrm{either}&\vert f_1\rangle\\
                            \mathrm{or}       &\vert f_2\rangle\end{array}
\end{equation}
 with the decay
time $\tau_G\sim\hbar/\ell^2_G$. This is the central postulate in the 
gravity-related (G-related) spontaneous collapse model, also called DP-model
after its proponents \cite{Dio86,Dio87,Dio89,Pen94,Pen96,Pen98,Pen04}.
From the above postulated collapse, it follows
that the pure Cat state becomes the mixture of its two componenets:
\begin{equation}
\frac{\vert f_1\rangle+\vert f_2\rangle}{\sqrt{2}}\frac{\langle f_1\vert+\langle f_2\vert}{\sqrt{2}}\Longrightarrow
 \frac{\vert f_1\rangle\langle{f_1}\vert+\vert f_2\rangle\langle{f_2}\vert}{2}.
\end{equation}
Accordingly, we talk about G-related spontaneous decoherence --- an intrinsically related mechanism 
to the spontaneous collapses. The dynamics of decoherence is simpler, being in terms of master equations for the density matrix
$\ro$ of the system in question; we take this approach in the forthcoming sections. 
(We shortly come back to the important distinction between collapse and decoherence in the Summary.) 

Since catness (\ref{catness}) would diverge for 
point-like constituents, a certain spatial cut-off $\si$ is needed.
The smaller the cut-off $\si$, the stronger will be the proposed spontaneous decay.
For a strong decay, one chooses $\si=10^{-12}$cm, to allow for a mass density
resolution as fine as the size of the nuclei \cite{Dio89}.  
Unfortunately, the detailed dynamics of the spontaneous collapses leads to
a constant rate of kinetic excitation for all microscopic constituents. 
This has been a basic problem, first pointed out in Ref.~\cite{GGR90}.

Traditionally, the DP-model used to be applied to single macroscopic d.o.f. like the c.o.m. of 
a bulk. The present work derives the DP-model for the hydrodynamic-elastic (acoustic)
d.o.f., opening new perspectives. 

\section{G-related decoherence}\label{G-RELATED_DECOHERENCE}
Let us start from a many-body system of Hamiltonian
\begin{equation}\label{Ham}
\Ho=\sum_a\frac{\pbh^2_a}{2m_a}+\sum_{a,b}V(\xbh_a-\xbh_b),
\end{equation}
where $m_a,\pbh_a,\xbh_a$ are the mass, and the canonical variables, respectively, of 
constituents. In the DP-model, the von Neumann evolution equation of the quantum state $\ro$ is
modified by the G-related decoherence term $\Dcal\ro$: 
\begin{equation}\label{ME}
\frac{d\ro}{dt}=-\frac{\im}{\hbar}[\Ho,\ro]+\Dcal\ro.
\end{equation}
This is the master equation of the DP-model
where $\Dcal=\Dcal\dg$ is proportional to the Newton gravitational constant $G$:
\begin{equation}\label{Dec}
\Dcal\ro=-\frac{G}{2\hbar}\int[\fh_\si(\rb),[\fh_\si(\rbp),\ro]]\frac{\dd\rb\dd\rbp}{\vert\rb-\rbp\vert}.
\end{equation}
The key quantity is the smoothened mass distribution operator: 
\begin{equation}\label{mdens}
\fh_\si(\rb)=\sum_a m_a g_\si(\rb-\xbh_a),
\end{equation} 
where $g_\si(\rb)$ is the central Gaussian distribution of width $\si$.
The finite width plays the role of cut-off since for point-like constituents
$\Dcal$ would diverge otherwise. We assume $\si\!\sim\!10^{-12}$cm which is about the 
nuclear size. In Fourier representation, using (\ref{mdens}), the decoherence (\ref{Dec}) takes this form:
\begin{equation}\label{DecFou}
\Dcal\ro=-\frac{G}{2\hbar}\int\frac{4\pi\erm^{-\kb^2\si^2}}{k^2}\sum_{a,b}m_a m_b[\erm^{\im\kb\xbh_a},[\erm^{-\im\kb\xbh_b},\ro]]\frac{\dd\kb}{(2\pi)^3}.
\end{equation}
Due to the decoherence term in (\ref{ME}), the total energy (\ref{Ham}) is not conserved. 
We can determine its Heisenberg time-derivative, yielding a number: 
\begin{equation}\label{dotH}
\frac{\dd\Ho}{\dd t}=\Dcal\Ho=\Dcal\sum_a\frac{\pbh^2_a}{2m_a}=\frac{1}{2\sqrt{4\pi}}G\hbar\si^{-3}M,
\end{equation}
where $M=\sum_a m_a$ is the total mass. For bulk matter, say for $M=1g$, the rate of spontaneous energy gain
is cca.$100$erg/s which would cause a gross eternal warming up, 
much too higher than in other collapse models     

Heating is an annoying feature of all spontananeous collapse models. 
From (\ref{dotH}) we realize that it is the kinetic energy of each constituent which
is increasing at a constant rate $\sim\!G\hbar m_a/\si^3$. To further characterize it, 
we define the  'nuclear' density
\begin{equation}\label{fanucl}
f^\nucl=\frac{m_\mathrm{av}}{(4\pi\si^2)^{3/2}}
\end{equation}
where $m_\mathrm{av}$ is the average constituent mass.
Let us consider the classical (non-quantum) frequency 
\begin{equation}\label{NewOsc}
\omG^\nucl=\sqrt{4\pi G f^\nucl/3}\sim 1 \mathrm{kHz}
\end{equation}
of the `Newton oscillator' \cite{Dio13a} in 'nuclear' density.
Now we can write the rate of spontaneous energy increase per microscopic d.o.f. as 
\begin{equation}\label{heat}
\frac{1}{2}\hbar(\omG^\nucl)^2\sim10^{-21}\mbox{erg/s}. 
\end{equation}
This is an extreme small value, but for an Avogadro number of constituents it
yields too much heating \cite{GGR90}, like $100$erg/s, as we said above.

\section{Acoustic mode decoherence}
To make Schr\"odinger Cats decay, which the DP-master equation (\ref{ME}) is good for,
the G-related spontaneous collapses 
of the macroscopic d.o.f .  matter. For macroscopic degrees of
freedom it is plausible to take the hydrodynamic-elastic (acoustic) ones.
In close-to-equilibrium states they decouple from the microscopic degrees
of freedom, therefore the dynamics of acoustic d.o.f. becomes
autonomous, the corresponding effective quantum state $\ro$ satisfies
a closed evolution equation with an effective Hamiltonian $\Ho$.
As we shall see, also the DP-decoherence term (\ref{DecFou}) induces a closed
form for the acoustic d.o.f. 

First, let us define the Hamiltonian part of the dynamics for a homogeneous bulk of mass $M$, 
volume $V$, and mass density 
\begin{equation}
\fa=\frac{M}{V}.
\end{equation} 
We start form the notion of displacement field known, e.g.,
from the theory of elasticity \cite{LanLif70}. We introduce the quantized displacement field $\ubh(\rb)$ together with the canonically
conjugated momentum field $\pibh(\rb)$, satisfying  the canonical commutators:
\begin{align}
[\hat{u}_i(\rb),\hat{u}_j(\rbp)]&=0,\nonumber\\
[\hat\pi_i(\rb),\hat\pi_j(\rbp)]&=0,~~~~~~~~~~~~~~~~~~(i,j=1,2,3)\nonumber\\
[\hat{u}_i(\rb),\hat\pi_j(\rbp)]&=\im\hbar\delta_{ij}\delta(\rb-\rbp).
\end{align}
 We assume that the macroscopic excitations of our bulk are quantized acoustic (sound) waves.
 For long wavelengths, they satisfy linear dynamics
 with the following Hamiltonian:
\begin{equation}\label{HamAc}
\Ho=\int\left(\frac{1}{2\fa}\pibh^2+\frac{\fa}{2}c_\ell^2(\mathbf{\nabla}\ubh)^2\right)\dd\rb,
\end{equation}
where $c_\ell$ is the longitudinal sound velocity. For simplicity's, we have restricted the calculations to the longitudinal 
modes satisfying $\mathbf{\nabla}\times\ubh=0$.

Second, let us determine the G-related spontaneous decoherence of the acoustic modes.
To this end, we re-express the decoherence  $\Dcal$ (\ref{Dec}) 
in function of the displacement field $\ubh(\rb)$.
We disregard the electronic constituents because of their small mass. 
We write the coordinate operators of the nuclei into this form:
\begin{equation}
\xbh_a=\xba_a+\ubh(\xba_a),
\end{equation}
where $\xba_a$ are the fiducial positions. 
If, furthermore, we assume that the displacements $\ubh(\rb)$ are much smaller than $\si$ then
in $\Dcal$ the cross-terms between different nuclei can be ignored and, in Fourier representation (\ref{DecFou}), 
the Taylor expansion $\exp[\im\kb\ubh(\xba_a)]\approx1+\im\kb\ubh(\xba_a)$ applies:
\begin{align}\label{DecXXX}
\Dcal\ro=&-\frac{G}{2\hbar}\int\frac{4\pi\erm^{-\kb^2\si^2}}{k^2}\sum_a m^2_a [\kb\ubh(\xba_a),[\kb\ubh(\xba_a),\ro]]\frac{\dd\kb}{(2\pi)^3}\nonumber\\
        =&-\frac{G}{2\hbar}\frac{1}{3\sqrt{4\pi\si^2}}\sum_a m^2_a [\ubh(\xba_a),[\ubh(\xba_a),\ro]]\nonumber\\
        \approx&-\frac{G}{2\hbar}\frac{1}{3\sqrt{4\pi}}m^2_\mathrm{av}\frac{\fa}{m_\mathrm{av}}\int[\ubh(\rb),[\ubh(\rb),\ro]]\dd\rb.
\end{align} 
The symbol $m^2_\mathrm{av}$ stands for the average squared mass of the nuclei. Let us define the 'nuclear' density as
\begin{equation}\label{fanuclHE}
f^\nucl=\frac{m^2_\mathrm{av}/m_\mathrm{av}}{(4\pi\si^2)^{3/2}},
\end{equation}
slightly different from (\ref{fanucl}), the same order of magnitude though. Similar will be the corresponding
Newton oscillator frequency $\omG^\nucl$ (\ref{NewOsc}).
Using it, we obtain the final form of the G-related decoherence term of the acoustic modes:
\begin{equation}\label{DecAc}
\Dcal\ro=-\frac{1}{2\hbar}\fa(\omG^\nucl)^2\int[\ubh(\rb),[\ubh(\rb),\ro]]\dd\rb.
\end{equation}

According to Eqs. (\ref{ME},\ref{HamAc},\ref{DecAc}), the decoherence master equation of the acoustic modes reads 
\begin{equation}\label{MEAc}
\frac{\dd\ro}{\dd t}=\frac{1}{2\hbar}
\int\left(\frac{-\im}{\fa}[\pibh^2,\ro]-\im\fa c_\ell^2[(\mathbf{\nabla}\ubh)^2,\ro]
-\fa(\omG^\nucl)^2[\ubh,[\ubh,\ro]]\right)\dd\rb.
\end{equation}
Recall that $\ubh(\rb),\pibh(\rb)$ are effective canonical variables of the long wavelength
acoustic modes. This feature will be elucidated in Fourier representation. 

\subsection{Fourier representation}\label{A}
Let us expand the canonical variables in terms of discrete Fourier components 
$\ubh_\kb=\ubh_{-\kb}\dg$ and $\pibh_\kb=\pibh_{-\kb}\dg$:
\begin{align}\label{canvarFou}
\ubh(\rb)&=\frac{1}{\sqrt{V}}\sum_\kb\ubh_\kb\erm^{\im\kb\rb},\nonumber\\ 
\pibh(\rb)&=\frac{1}{\sqrt{V}}\sum_\kb\pibh_\kb\erm^{\im\kb\rb}.  
\end{align}
satisfying the discrete canonical commutation relationships
\begin{align}\label{canrelFou}
[\hat{u  }_{\kb i},\hat{u  }_{\lb j}\dg]&=0,\nonumber\\
[\hat{\pi}_{\kb i},\hat{\pi}_{\lb j}\dg]&=0,~~~~~~~~~~~~~~~~~~(i,j=1,2,3)\nonumber\\
[\hat{u  }_{\kb i},\hat{\pi}_{\lb j}\dg]&=\im\hbar\delta_{ij}\delta_{\kb\lb}.
\end{align}
The Hamiltonian (\ref{HamAc}) and decoherence (\ref{DecAc}) read, respectively:
\begin{align}\label{HamDecFou}
\Ho&=\frac{1}{2}\sum_\kb
\left(\frac{1}{\fa}\pibh_\kb\dg\pibh_\kb+\fa c_\ell^2k^2\ubh_\kb\dg\ubh_\kb\right),\\
\Dcal\ro&=-\frac{1}{2\hbar}\sum_\kb\fa(\omG^\nucl)^2[\ubh_\kb\dg,[\ubh_\kb,\ro]].
\end{align}
The master equation (\ref{MEAc}) takes the following form for the acoustic Fourier modes:
\begin{equation}\label{MEAcFou}
\frac{\dd\ro}{\dd t}=\frac{1}{2\hbar}\sum_\kb
\left(\frac{-\im}{\fa}[\pibh_\kb\dg\pibh_\kb,\ro]-\im\fa c_\ell^2k^2[\ubh_\kb\dg\ubh_\kb,\ro]
-\fa(\omG^\nucl)^2[\ubh_\kb\dg,[\ubh_\kb,\ro]]\right).
\end{equation}
Now we can calculate the heating rate:
\begin{equation}\label{dotHac}
\frac{\dd\Ho}{\dd t}=\Dcal\Ho=\sum_\kb\Dcal\frac{\pibh_\kb\dg\pibh_\kb}{2\fa}=\sum_\kb\frac{3}{2}\hbar(\omG^\nucl)^2.
\end{equation}
Observe that each acoustic mode undergoes the same tiny heating similar to the heating (\ref{heat}) found for the individual d.o.f. of 
each constituent.

\subsection{Center of mass decoherence}\label{B}
The c.o.m. motion of the bulk is decoupled from the internal acoustic modes. Let us
read out the dynamics of the c.o.m. position $\Xbh$ and momentum $\Pbh$ from (\ref{canvarFou}):
\begin{align}\label{canvarcom}
\Xbh&=\frac{1}{\sqrt{V}}\ubh_\mathbf{0},\nonumber\\ 
\Pbh&=\sqrt{V}\pibh_\mathbf{0},  
\end{align}
where we set the fiducial c.o.m. position to the origin.
We identify the c.o.m. parts of the master equation (\ref{MEAcFou}):
the free body kinetic Hamiltonian $\Pbh^2/2M$ and the standard position decoherence
\begin{equation}
\Dcal_\mathrm{c.o.m.}\ro=-\frac{1}{2} M(\omG^\nucl)^2[\Xbh,[\Xbh,\ro]].
\end{equation}
The c.o.m. dynamics is thus governed by the autonomous master equation
\begin{equation}
\frac{\dd\ro_\mathrm{c.o.m.}}{\dd t}=
\frac{-\im}{\hbar}\left[\frac{\Pbh^2}{2M},\ro_\mathrm{c.o.m.}\right]-\frac{1}{2\hbar}M(\omG^\nucl)^2[\Xbh,[\Xbh,\ro_\mathrm{c.o.m.}]],
\end{equation}
in full accordance with the old derivations in the DP-model \cite{Dio86,Dio87,Dio89}.
If we calculate the heating rate we get
\begin{equation}
\Dcal_\mathrm{c.o.m.}\frac{\Pbh^2}{2M}=-\frac{1}{4\hbar}(\omG^\nucl)^2[\Xbh,[\Xbh,\Pbh^2]]=\frac{3}{2}\hbar(\omG^\nucl)^2\sim10^{-21}\mathrm{erg/s}.
\end{equation}
This is the same extreme small value (\ref{heat}) that we obtained universally for each individual constituent or, alternatively, for each acoustic mode (\ref{dotHac}).

\subsection{Universal dominance of decoherence}\label{C}
In bulk matter, the DP-model yields a certain simple universal
behaviour of spontaneous decoherence.  In the master equation  (\ref{MEAcFou}), 
consider the magnitudes of the harmonic potential and
decoherence terms, respectively. Both of them are
quadratic in the displacements $\ubh_\kb$. Although their structure is different, we see that the
harmonic potential becomes suppressed by the decoherence term for small wave numbers $k$ such that
\begin{equation}
c_\ell k \ll \omG^\nucl.
\end{equation}
In solids, e.g., the typical range of sound velocity is $c_\ell\sim10^5$cm/s, the above condition means
wavelengths larger than $\sim\!1$m. The master equation (\ref{MEAcFou}) for these modes
takes the following form:
\begin{equation}
\frac{\dd\ro}{\dd t}=
\frac{1}{2\hbar}\!\!\sum_{1\!/\!k\gg 1m}
\left(\frac{-\im}{\fa}[\pibh_\kb\dg\pibh_\kb,\ro]
-\fa(\omG^\nucl)^2[\ubh_\kb\dg,[\ubh_\kb,\ro]]\right).
\end{equation}
In oscillatory modes of wavelength $\gg\!\!1$m, the G-related decoherence dominates over the directional force.
Suppose we have a bulk of rock as big as $100$m. Consider a
sub-volume inside, with size about a few meters at least. Then the c.o.m. of this inside body behaves as if
the body were a free-body subject to c.o.m. spontaneous decoherence, like in Sec. \ref{B}.
The directional force from the behalf of the environmental rock is not absent, of course. On a time
scale much longer than spontaneous decoherence's, it will keep the inside body close to its fiducial position. 

\subsection{Strong spontaneous decoherence at low heating}\label{D}
Consider the master equation (\ref{MEAcFou}) of the DP-model for the acoustic modes. As we said in Sec. \ref{A},
each mode undergoes the heating rate (\ref{heat}). 
Is it possible, by some refinement of the DP-model, to reduce the
spontaneous heating but to retain the strength of decoherence in the macroscopic d.o.f.? 

Let us choose a larger cut-off $\si$, say hundred times the 'nuclear' size.
The ominous parameter $f^\nucl$ (\ref{fanucl},\ref{fanuclHE}) would drop by six orders of magnitude,
 resulting in six orders of magnitude reduction of heating rate at the price of the same reduction of the strength of spontaneous 
collapses. The critical size, $\sim\!1$m in Sec. \ref{C}, where c.o.m. DP-collapses become faster than the directional forces,
will increase by six orders of magnitude. So we cannot play much with the cut-off $\si$.

Instead, we can play with the number of acoustic modes. Suppose that short wave acoustic
modes are not subjected to G-related spontaneous collapses. For instance, let us set
this limit to $\lambda=10^{-5}$cm, i.e., we replace the standard master equation (\ref{MEAcFou})
by the following version:
\begin{equation}\label{MEAcFou1}
\frac{\dd\ro}{\dd t}=\frac{-\im}{2\hbar}\sum_\kb
\left(\frac{1}{\fa}[\pibh_\kb\dg\pibh_\kb,\ro]+\fa c_\ell^2k^2[\ubh_\kb\dg\ubh_\kb,\ro]\right)
-\frac{1}{2\hbar}\sum_{1\!/\!k\gg\lambda}\fa(\omG^\nucl)^2[\ubh_\kb\dg,[\ubh_\kb,\ro]].
\end{equation}
Since $\lambda$  is three orders of magnitude larger than the internuclear
distance in common bulk matter, the number of spontaneously heated acoustic
modes drops by nine orders of magnitude compared to the number of the nuclei. 
The spontaneous heating rate of $1$g will reduce to $10^{-7}$erg/s instead of $100$erg/s
found in Sec. \ref{G-RELATED_DECOHERENCE}.  
The strength and dynamics of G-related spontaneous collapses of the long wavelengths 
acoustic modes, including the c.o.m. as well, remain the same as before.

\section{Summary}
We have derived the DP-model of G-related spontaneous collapses for the hydrodynamic-elastic (acoustic)
d.o.f. of bulk matter. To assure strong significance of collapses, we chose the minimum plausible
cut-off $\si$ which is about the nuclear size. This leads to the dominance of the spontaneous
collapses over the elastic forces inside common condensed matter for wavelengths larger than about $1$m.
The warming up, an annoying side-effect of spontaneous collapses, will considerably drop if we 
ascribe spontaneous collapses to the really macroscopic acoustic modes only. This modification does not
influence the usual predictions of the model concerning the collapse in macroscopic d.o.f. 

For simplicity's sake, we worked out the master equations of spontaneous decoherence of the acoustic modes. 
We spared the now straightforward derivation of the stochastic (jump \cite{Dio86} or diffusive \cite{Dio89})
Schr\"odinger equations of G-related spontaneous collapses for the acoustic modes. As we mentioned in the Introduction,
spontaneous decoherence and collapse are to be distinguished conceptionally. Spontaneous decoherence is the 
testable local effect of spontaneous collapse. It can be mimicked by and it is usually masked by environmental 
decoherence. Continued laboratory efforts are trying to suppress these environmental effects 
\cite{Mar_Bou03,Van_Asp11,Rom11,LiKheRai11,Pep_Bou12,Quietal13,YinGerLi13,JufUlbArn13}. 
On the contrary, collapse is global effect, cannot be mimicked or masked by the environment however noisy
it is \cite{Dio13a}. In any current models of spontaneous collapse \cite{Basetal13}, collapse itself is 
never detectable, only the resulting spontaneous decoherence is, as emphasized in Ref. \cite{Dio14a}. 
To let spontaneous collapses be testable, recent extension of the DP-model has shown interesting theoretical 
and experimental perspectives \cite{Dio13a,Dio13b,Dio14a,Dio14b}.  

On the spontaneous decoherence of the acoustic modes, derived in the present work, we remark that it
might be influenced or even masked by the modes' higher order coupling to the microscopic d.o.f. inside 
the bulk or from the behalf of the environment. Further related investigations are certainly needed.

This work was supported by the Hungarian Scientific
Research Fund (Grant No. 75129) and the EU COST
Action MP1006.


\begin{thebibliography}{9}
\bibitem{Dio86} Di\'osi L 1986 {\it A quantum-stochastic gravitation model and the reduction of the wavefunction} 
                (Thesis, in Hungarian) \\ http://www.rmki.kfki.hu/$\!\sim$diosi/thesis1986.pdf
\bibitem{Dio87} Di\'osi L 1987 {\it Phys. Lett.} {\bf 120 A} 377
\bibitem{Dio89} Di\'osi L 1989 {\it Phys. Rev.} {\bf A 40} 1165
\bibitem{Pen94} Penrose R 1994 {\it Shadows of the mind} (Oxford: Oxford University Press)
\bibitem{Pen96} Penrose R 1996 {\it Gen. Rel. Grav.} {\bf 28} 581
\bibitem{Pen98} Penrose R 1998 {\it Phil. Trans. R. Soc. Lond.} {\bf A 356} 1927
\bibitem{Pen04} Penrose R 2004 {\it The Road to Reality} (London: Jonathan Cape Publishers)
\bibitem{GGR90} Ghirardi G C, Grassi R, Rimini A 1990 {\it Phys. Rev.} {\bf A 42} 1057 
\bibitem{Dio13a} Di\'osi L 2013 {\it J. Phys. Conf. Ser.} {\bf 442} 012001
\bibitem{LanLif70} Landau L D,  Lifshitz E M 1970 {\it Theory of Elasticity} (Oxford: Pergamon Press) 
\bibitem{Mar_Bou03} Marshall W, Simon C, Penrose R, Bouwmeester D 2003 {\it Phys. Rev. Lett.} {\bf 91} 130401
\bibitem{Van_Asp11} Vanner M R, Pikovski I, Cole G D, Kim M S, Brukner \v{C}, Hammerer K, Milburn G J, Aspelmeyer M
                    2011 {\it PNAS} {\bf 108} 16182
\bibitem{Rom11} Romero-Isart O 2011 {\it Phys. Rev.} {\bf A 84} 052121
\bibitem{LiKheRai11} Li T, Kheifets S, Raizen M G 2011 {\it Nat. Phys.} {\bf 7} 527
\bibitem{Pep_Bou12} Pepper B, Ghobadi R, Jeffrey E, Simon C, Bouwmeester D 2012 {\it Phys. Rev. Lett.} {\bf 109} 023601
\bibitem{Quietal13} Quinn T, Parks H, Speake C, Davis R 2013 {\it Phys. Rev. Lett.} {\bf 111} 101102
\bibitem{YinGerLi13} Yin Z Q, Geraci A A, Li T 2013 {\it Int. J. Mod. Phys.} {\bf B 27} 1330018
\bibitem{JufUlbArn13} Juffmann T, Ulbricht H, Arndt M 2013 {\it Rep. Prog. Phys.} {\bf 76} 086402
\bibitem{Basetal13} Bassi A, Lochan K, Satin S,  Singh T P, Ulbricht H 2013  {\it Rev. Mod. Phys.} {\bf 85} 471
\bibitem{Dio14a} Di\'osi L 2014 {\it J. Phys. Conf. Ser.} {\bf 504} 012020
\bibitem{Dio13b} Di\'osi L 2013 {\it Phys. Lett.} {\bf A 377} 1782
\bibitem{Dio14b} Di\'osi L 2014 {\it Found. Phys.} {\bf 44} 1

\end{thebibliography}
\end{document}